\documentclass{article}\usepackage{amsmath,amssymb,textcomp}
\usepackage{graphicx,mathrsfs}

\begin{document}

\title{Can Quantum Markov Evolutions Ever Be Dynamically Decoupled?}

\author{John E. Gough\thanks{J. E. Gough is with the Department of Physics, Aberystwyth University, UK. (\texttt{email: jug@aber.ac.uk}).}\and Hendra I. Nurdin\thanks{H. I. Nurdin is with the School of Electrical Engineering and 
Telecommunications,  UNSW Australia,  Sydney NSW 2052, Australia (\texttt{email: h.nurdin@unsw.edu.au}). 
Research supported by the Australian Research Council.}   }


\maketitle

\begin{abstract}
We consider the class of quantum stochastic evolutions ($SLH$-models) leading to a quantum dynamical semigroup over a fixed quantum mechanical system (taken to be finite-dimensional). We show that if the semigroup is dissipative, that is, the coupling operators are non-zero, then a dynamical decoupling scheme based on unitary rotations on the system space cannot suppress decoherence even in the limit where the period between pulses vanishes. We emphasize the role of the Fock space dilation used here to construct a quantum stochastic model, as there are often dilations of the same semigroup using an environmental noise model of lower level of chaoticity for which dynamical decoupling is effective. We show that the Chebotarev-Gregoratti Hamiltonian behind a quantum stochastic evolution
is an example of a Hamiltonian dynamics on a joint system-environment that cannot be dynamically decoupled in this way.
\end{abstract}

\section{Introduction}
Dynamical decoupling (DD) was introduced by Viola and Lloyd \cite{VL98} as a proposal to suppress decoherence effects in quantum open systems by applying rapid periodic \lq\lq unitary kicks\rq\rq , drawn from a fixed set of unitaries: the desired effect is that the noisy dynamical effects on the system, due to coupling with its environment, get averaged to zero up to a certain order with respect to period of the kicks \cite{VL98}-\cite{TV06}. It has since developed into an important open-loop approach for decoherence and error suppression \cite{KL05}-\cite{K_13}. For a review of
the development and state-of-the-art see the recent review article by Lidar \cite{Lidar}. For recent discussions on the Markovianity versus non-Markovianity see
 \cite{SA15} and \cite{ACCMP15}

Our interest will be in the question of whether the approach works for open systems. In particular, suppose we are given a fixed quantum dynamical semigroup (or equivalently, a master equation) can we say whether a prescribed DD scheme will be effective.
The question turns out to be misstated - a quantum dynamical semigroup may have several inequivalent dilations, some of which are amenable to dynamical decoupling and some of which are not. The correct question is whether a given dilation is amenable to dynamical decoupling. Physically we have no choice but to work with the actual environment, and the restriction is that the decoupling scheme can only be applied to the system (not the environment). We shall focus on environments which allow a quantum stochastic description - so called $SLH$-models
\cite{CKS}.

\subsection{Dynamical Decoupling}
We fix a quantum mechanical system with a finite-dimensional Hilbert space $\mathfrak{h}_{\text{sys}}=\mathbb{C}^{d}$. Suppose we have a unitary dynamics given by a family $U\left( t,s\right) $ of unitaries, $t\geq s$, satisfying the flow property 
\begin{eqnarray}
U\left( t,s\right) =U\left( t,r\right) U\left( r,s\right) ,  \label{eq:flow}
\end{eqnarray}
for $t\geq r\geq s$. For instance, suppose we have a time dependent Hamiltonian $H(t)$, then we may take $U(t,s)$ to be the solution to the Schr\"{o}dinger equation $\frac{\partial }{\partial t} U(t,s) = -i H(t) \, U(t,s)$ for $ t \ge s$ with $U(t=s , s)=I$. Formally, we may write
\begin{eqnarray}
U\left( t,s\right) = \vec{T} e^{-i \int_t^s H( \tau ) d \tau},  \label{eq:TOE}
\end{eqnarray}
where $\vec{T}$ denotes chronological ordering. (Note that we will encounter open system models latter where the system is interacting with an environment, but where the same flow equation (\ref{eq:flow}) arises. In this case we will have a
quantum stochastic description for the environment, rather than the type of Schr\"{o}dinger equation consider at present.)

Suppose now that we have a finite-dimensional system coupled to a bath through the Hamiltonian $H=H_{\rm sys} + H_{\rm SB} + H_{\rm B}$, where $H_{\rm sys}$ is the system Hamiltonian, $H_{\rm SB}$ is the interaction Hamiltonian between the system and bath, and $H_{\rm B}$ is the bath Hamiltonian. We are interested in the case of preserving the identity operation as in a quantum memory, so we wish to eliminate the effect of $H$ on the {\em system only} and maintain that $U(t,0) \approx I \otimes Z(t)$ for  some (time-varying) bath operator $Z(t)$. So, the interaction can affect the bath in a non-trivial way but we are not concerned at all about this. In this case $U(t,s) =  \exp(-\imath H(t-s))$ and $\{ U(t,0)\}$ is a unitary group. A dynamical decoupling (DD) scheme \cite{VL98,VKL99,VK03} is a modification to the dynamics where, over a specific time interval $\left[ s,t\right] $, we apply
unitaries chosen from a finite set $\mathscr{V}$ at times $t=t_{N}\geq t_{N-1}\geq \cdots \geq t_{0}=s$ so that the unitary evolution becomes 
\begin{eqnarray}
\tilde{U}_{\mathbf{v}}\left( t,t_0\right) =\tilde{U}_{v_{N}}\left(
t_{N},t_{N-1}\right) \cdots \tilde{U}_{v_{2}}\left( t_{2},t_{1}\right) \tilde{U}_{v_{1}}
\left( t_{1},t_{0}\right)  \label{eq:U_SQT}
\end{eqnarray}
where 
\begin{eqnarray}
\tilde{U}_{v_{k}}\left( t_{k+1},t_{k}\right) =v_{k}^{\ast }U\left(
t_{k+1},t_{k}\right) v_{k}  \label{eq:SQT_DD_step}
\end{eqnarray}
and $\mathbf{v}=\left( v_{N},\cdots ,v_{1}\right) $ is a sequence drawn from $\mathscr{V}$. A common strategy is to fix the set $\mathscr{V}$ so that, for any operator $X$, we have
\begin{eqnarray}
\frac{1}{\#\mathscr{V}}\sum_{v\in \mathscr{V}}v^{\ast }Xv=\frac{1}{d}\mathrm{%
tr} \big\{ X\,I_{\text{sys}} \big\}.  \label{eq:DD_identity}
\end{eqnarray}
For instance, if $\mathscr{V}$ is a group then the group average is a conditional expectation onto the group centralizer, so it suffices to choose the group to have trivial centralizer. Such a group is called a {\em decoupling group}.  For the  quantum memory scenario described  above, applying a DD sequence based on a decoupling group one gets that $\tilde{U}_{\mathbf{v}}(t_N,t_0) = I \otimes Z(t) + O(t_N^2)$, where $N=\# \mathscr{V}$ and $Z(t)$ an arbitrary bath operator. So DD suppresses the effect of the Hamiltonian $H$ to first order in $t_N$.

We shall assume that for large $N$ the sequences $\mathbf{v}$ have the property that each element $v\in \mathscr{V}$ occurs uniformly. It is often sufficient to assume that the sequence cycles deterministically through the elements of $\mathscr{V}$ in some fixed order. Alternatively, we may assume that the $v_{k}$ form an i.i.d. sequence of random variables taking values in $\mathscr{V}$, with a common distribution being the uniform distribution.

For definiteness, we take $t_{k+1}-t_{k}=\tau $ to be a fixed time-step which we call the \textit{dynamical decoupling time step}, for each $k$. We make the following assumption: for about $N_{dd}$ steps we have that 
\begin{eqnarray}
\frac{1}{N_{dd}}\sum_{k=1}^{N_{dd}}v_{k}^{\ast }Xv_{k}\approx \frac{1}{\#%
\mathscr{V}}\sum_{v\in \mathscr{V}}v^{\ast }Xv.  \label{eq:LLN}
\end{eqnarray}
This is admittedly a rough and ready statement: for the case where we are cycling through elements of $\mathscr{V}$ then we need only take $N_{dd}=\# \mathscr{V}$ and obtain the right-hand side exactly by virtue of (\ref{eq:DD_identity}), however, for the randomized case we have in mind a law of large numbers result. In the latter case, we can of course make the statement mathematically precise. Note that (\ref{eq:DD_identity}) and (\ref{eq:LLN}) in combination gives 
\begin{eqnarray}
\frac{1}{N_{dd}}\sum_{k=1}^{N_{dd}}v_{k}^{\ast }Xv_{k}\approx \frac{1}{d}\mathrm{tr}\{X\}\,I_{\text{sys}}.
\end{eqnarray}

Our main point however is that there is a longer time-scale $T_{dd}=N_{dd}\tau $ at which the averaging takes place and dynamical
decoupling starts to become effective.

\section{Open Quantum Systems}

A quantum dynamical semigroup, $\left( \Phi _{t}\right) _{t\geq 0}$, is a family of norm-continuous completely positive maps on the algebra of bounded operators on a fixed Hilbert space $\mathfrak{h}_{\text{sys}}$, taken as $\mathbb{C}^{d}$, with $\Phi _{t}\circ \Phi _{s}=\Phi _{t+s}$. As is well known, such semigroups possess a generator of the form
\begin{eqnarray}
\mathcal{L}\left( X\right) =\frac{1}{2}\sum_{k}\left[ L_{k}^{\ast },X\right]
L_{k}+\frac{1}{2}\sum_{k}L_{k}^{\ast }[X,L_{k}]-i\left[ X,H\right] 
\label{eq:GKS_Lindblad}
\end{eqnarray}
known as a GKS-Lindblad generator \cite{GKS}-\cite{ALV}, so that we may write $\Phi _{t}=e^{t\mathcal{L}}$. (Here $H$ will be self-adjoint bounded, and the $L_k$ are bounded.) Given a density matrix $\varrho _{\text{sys}}$ for the system, we obtain an average evolution
\begin{eqnarray}
w_t (X) \triangleq \text{tr}_{\text{sys}}\left\{ \varrho _{\text{sys}}\,\Phi _{t}\left( X\right) \right\} .
\end{eqnarray}

However, this is just the first level in a hierarchy of expectations.
The \textit{Markovian time-ordered correlation kernels} of the semigroup are defined to be
\begin{gather}
 w_{t_1 , \cdots , t_n} ( Y_1 , \cdots , Y_n ; X_1 , \cdots , X_n) \triangleq  \nonumber \\
 \text{tr}_{\text{sys}} \bigg\{ \varrho _{\text{sys}} \Phi_{ \tau_1} \bigg( Y_1^\ast \Phi_{\tau_2 } \big( Y^\ast_{2} \cdots 
\Phi_{\tau_n } (Y^\ast_n X_n) \cdots X_2 \big) X_1 \bigg) \bigg\}, \qquad
\label{eq:multitime}
\end{gather}
where $0 \leq t_1 \leq t_2 \leq \cdots \leq t_n$, and we have $\tau_1 = t_1$, and $ \tau_k = t_{k} - t_{k-1}$ ($k=2, \cdots, n$).
This is known as the \lq\lq quantum regression\rq\rq\, formula in the physics literature \cite{Markov}. The rationale for considering only multi-time correlations with these specific time-orderings is well known in quantum physics - see for instance \cite[Section 2.3]{Gardiner}. 

A central question is whether we can dilate the semigroup. Here we mean having family of maps $j_t (\cdot )$ from the algebra of operators on $\mathfrak{h}_{\text{sys}}$ into a \lq\lq larger \rq\rq\, algebra $\mathfrak{A}$ with a faithful normal 
state $\mathbb{E}$ on the algebra such that the time-ordered correlations were given by
\begin{eqnarray}
\mathbb{E} \big[ j_{t_1} (Y_1^\ast ) \cdots j_{t_n} ( Y_n^\ast) j_{t_n } ( X_n)  \cdots j_{t_1} (X_n) \big] .
\end{eqnarray}
The dilation is termed \textit{unitary} if we also have
\begin{eqnarray}
j_t (X) \equiv V(t)^\ast j_0 (X) V(t)
\end{eqnarray}
where $(V(t))_{t \in \mathbb{R}}$ is a unitary group.

Here we encounter a problem similar to the Kolmogorov reconstruction theorem for a stochastic process from its finite-dimensional distributions.
The quantum stochastic process can be reconstructed uniquely from its general correlation kernels, but this is a larger family that just the time-ordered ones \cite{AFL}. As such one will generally have a choice of inequivalent unitary dilations (for instance, the large algebra may the tensor product of the system algebra with that of either a Boson or Fermion bath). Supplementary conditions to fix the dilation include stationarity and detailed balance \cite{FG84}.

The dissipation associated with a super-operator $\mathcal{L}$ is defined to be
\begin{equation*}
D_{\mathcal{L}}\left( X,Y\right) =\mathcal{L}\left( X^{\ast }Y\right) -
\mathcal{L}\left( X^{\ast }\right) Y-X^{\ast }\mathcal{L}\left( Y\right) ,
\end{equation*}
and one of the key properties of Lindblad generators is that they are dissipative, that is $D_{\mathcal{L}}\left( X,X\right) \geq 0$ for all $X$. (In fact, they must be completely dissipative \cite{Lindblad}.) Explicitly, for the form (\ref{eq:GKS_Lindblad}), we have $D_{\mathcal{L}}\left( X,X\right) =\sum_{k}\left[ X,L_{k}\right] ^{\ast }\left[ X,L_{k}\right] $ and we see that the Hamiltonian part of 
(\ref{eq:GKS_Lindblad}), that is, the super-operator $-i\left[ \cdot ,H\right] $, makes no contribution to the dissipation. The remaining part, $\mathcal{L}_{\textrm{diss}} \equiv \frac{1}{2}\sum_{k}\left\{ L_{k}^{\ast }\left[ \cdot ,L_{k}\right] +\left[ L_{k}^{\ast
},\cdot \right] L_{k}\right\} $ is called the dissipative part and is
determined by the dissipation $D_{\mathcal{L}}\left( \cdot ,\cdot \right) $.

\subsection{A Minimal Stochastic Model}

Let us take our system to have a Hamiltonian $H_{\text{sys}}=\sum_{n}E_{n}P_{n}$, where $P_{n}$ are orthonormal projections with 
$\sum_{n}P_{n}=I_{\text{sys}}$. The closed system dynamics under $H_{\text{sys}}$ leads to
\begin{eqnarray*}
w^{\text{free}}_t(X)  &\equiv &\text{tr}\left\{
\varrho _{\text{sys}}\,\,e^{itH_{\text{sys}}}Xe^{-itH_{\text{sys}}}\right\} 
\\
&\equiv &\sum_{n,m}\text{tr}\left\{ \varrho _{\text{sys}}\,\,P_{n}XP_{m}%
\right\} e^{-i\left( E_{m}-E_{n}\right) t}.
\end{eqnarray*}

We now randomize the model in a very simple way: we rescale the Hamiltonian as
\begin{eqnarray}
H(t, \Lambda) = \Lambda \, H_{\text{sys}}, 
\end{eqnarray}
where $\Lambda$ is a real-valued random variable having a Cauchy distribution with the probability density function 
$\rho ( \lambda ) =\frac{1}{\pi }\frac{1}{\lambda ^{2}+1}$. If the value of $\Lambda$ is known to be $\lambda$ then it is clear that the average value of an observable $X$ at time $t$ should be $ w^{\text{free}}_{\lambda t}(X)$, so if we are sampling from this ensemble of models we find the average
\begin{eqnarray*}
\mathbb{E}_t [ X ] = \int_{-\infty }^{\infty } w^{\text{free}}_{\lambda t}( X) \, \rho (\lambda ) d\lambda ,
\end{eqnarray*}
and, using the fact that 
 $\int_{-\infty }^{\infty }e^{iu\lambda }\, \rho (\lambda ) d\lambda =e^{-|u|}$ for the Cauchy distribution, we find that this equals
\begin{eqnarray}
\mathbb{E}_t [ X ] =\sum_{n,m}\text{tr}\left\{ \varrho _{\text{sys}}\,\,P_{n}XP_{m}%
\right\} e^{- | E_{m}-E_{n}| \,  t}.
\label{eq:E_t}
\end{eqnarray}
We see that we may write $\mathbb{E}_t [ X ]$ as 
\begin{eqnarray}
w_t (X) \equiv  \text{tr} \big\{ \varrho_{\text{sys}} \, \Phi _{t}\left( X\right)\big\}  ,
\label{eq:first}
\end{eqnarray}
where
\begin{eqnarray}
\Phi _{t}\left( X\right) =\sum_{n,m}   P_{n}XP_{m} \, e^{-|E_{m}-E_{n}|\, t}.
\label{eq:Phi_BW}
\end{eqnarray}
By inspection, $\Phi _{t}$ defines a semigroup and by construction (as a partial trace) is completely positive - therefore it forms a quantum dynamical semigroup with GKS-Lindblad generator
\begin{eqnarray*}
\mathcal{L}X=-\sum_{n,m}|E_{m}-E_{n}|\,P_{n}XP_{m}.
\label{eq:BW_GKSL}
\end{eqnarray*}

We note that instead of a classical Cauchy distributed random variable $\Lambda$, we could have used a more elaborate quantum model based on the Breit-Wigner description of an \lq\lq unstable particle\rq\rq . We take the unstable particle Hilbert space to be $L^{2}\left( \mathbb{R}\right) $ (the space of all complex square integrable functions on $\mathbb{R}$) and fix the self-adjoint operator $\left( H_{\text{B.W.}}f\right) \left( \lambda \right) =\lambda f\left( \lambda \right) $. Let $\varphi _{\text{B.W.}}$ be the wavefunction
\begin{eqnarray*}
\varphi _{\text{B.W.}}\left( \lambda \right) =\frac{1}{\sqrt{\pi }}\frac{1}{%
\lambda +i}.
\end{eqnarray*}
Note that the density $|\varphi _{\text{B.W.}}\left( \lambda \right) |^{2}$ is the desired Cauchy distribution (Lorentzian spectral density in physics language). The dilation is then given by taking the large algebra $\mathfrak{A}$ to be the algebra of bounded operators on the system tensored with the algebra of bounded operators on the unstable particle. Here $j_0 (X) = X \otimes I_{\text{B.W.}}$ and we take the unitary group to be
\begin{eqnarray*}
V\left( t\right) =e^{-itH_{\text{sys}}\otimes H_{\text{B.W.}}}.
\end{eqnarray*}
The state of the large system is taken to be
\begin{eqnarray*}
\varrho _{\text{sys}} \otimes |\varphi _{\text{B.W.}}\rangle \langle \varphi _{\text{B.W.}}|.
\end{eqnarray*}
In this case we find $\mathbb{E} [ j_t (X) ] = \mathbb{E}_t [X]$ as given in (\ref{eq:E_t}) above. Technically the quantization of the random variable $\Lambda$ is unnecessary, but allows us to frame the model in the language of quantum dilations.

An example of this is the \lq\lq shallow pocket\rq\rq\, model introduced by Arenz \textit{et al.} \cite{A_15} where they take
$\mathfrak{h}_{\text{sys}}= \mathbb{C}^{2}$ and $H_{\text{sys}}=\gamma \sigma _{z}$, the $z$ Pauli matrix: the resulting semigroup describes dephasing. For the choice $\mathscr{V}=\left\{ I,\sigma _{x}\right\} $, we have however that
\begin{eqnarray*}
\frac{1}{\# \mathscr{V}}\sum_{v\in \mathscr{V}}(v\otimes I_{\text{B.W.}})^{\ast }\left( H_{\text{sys}%
}\otimes H_{\text{B.W.}}\right) (v\otimes I_{\text{B.W.}})\equiv 0
\end{eqnarray*}
since $\sigma _{x}\sigma _{z}\sigma _{x}=-\sigma _{z}$. 
A general discussion of such models in which dynamical decoupling works is given in \cite{A_15}. 

Before leaving this model, we note that it is not stationary. In fact, we may explicitly compute the two-time kernels
\begin{eqnarray}
w_{t,t+h} (Y,X) &\equiv& \sum_{n,m,r} \text{tr} \big\{ \varrho_{\text{sys}} P_n Y^\ast P_m X P_r \big\} \nonumber\\
&& \times e^{- | (E_n - E_r) t+(E_m-E_r ) h |},
\label{eq:w_shallow}
\end{eqnarray}
which is clearly $t$-dependent. Furthermore, it is not Markovian since otherwise the two-point kernel from (\ref{eq:multitime}) would be given by
\begin{eqnarray}
w_{t,t+h} (Y,X) &=&\text{tr} \bigg\{ \varrho_{\text{sys}} \, \Phi_t \big( Y^\ast \Phi_h (X) \big) \bigg\} \nonumber\\
& \equiv&  \sum_{n,m,r} \text{tr} \big\{ \varrho_{\text{sys}} P_n Y^\ast P_m X P_r \big\}  \nonumber \\
&&\quad \times e^{- | E_n - E_r| t - |E_m-E_r | h },
\end{eqnarray}
which differs from (\ref{eq:w_shallow}) for $E_r$ intermediate between $E_n$ and $E_m$.

We see from (\ref{eq:first}) that these models give the correct one-point function for averages associated with 
the quantum dynamical semigroup in (\ref{eq:Phi_BW}), but the higher-order ($n>1$) time-ordered correlation kernels
arte not those of a Markov dilation.

\section{Quantum Markov Dynamics}
\label{sec:QSDE_models}
In this section, we turn to the natural class of dilation of quantum dynamical semigroups based on the Hudson-Parthasarathy quantum stochastic calculus \cite{HP,partha}. Here we have explicit constructions for the semigroup with given GKS-Lindblad generator $\mathcal{L}$ in (\ref{eq:GKS_Lindblad}).

We again take our system to have Hilbert space $\mathfrak{h}_{\text{sys}}=\mathbb{C}^{d}$, but now allow it to interact with an environment in a Markovian manner. The noise consists of $n$ quantum input processes which act on an auxiliary Hilbert space $\mathfrak{F}$ which is a Boson Fock space for the $n$ continuous variable processes: the noise will be taken to be in the Fock vacuum state $|\Omega \rangle $. Specifically, we consider a Hudson-Parthasarathy quantum stochastic evolution determined by the unitary family $U\left( t,s\right) \in \mathfrak{h}_{\text{sys}}\otimes \mathfrak{F}$ , $t\geq s$, satisfying the quantum stochastic differential equation \cite{HP}, \cite{partha}
\begin{eqnarray}
dU\left( t,s\right)  &=&\bigg\{ \sum_{j=1}^{n}L_{j}\otimes dB_{j}\left( t\right)
^{\ast }-\sum_{j=1}^{n}L_{j}^{\ast }\otimes dB_{j}\left( t\right)   \nonumber
\\
&&-\left( \sum_{j=1}^{n}L_{j}^{\ast }L_{j}+iH\right) dt \bigg\} U\left( t,s\right) ,
\label{eq:QSDE}
\end{eqnarray}
with $U\left( s,s\right) =I_{\text{sys}}\otimes I_{\mathfrak{F}}$. The family is an adapted unitary process and satisfies the flow property (\ref{eq:flow}). We also note that the non-vanishing component of the quantum It\={o} table is
\begin{eqnarray}
dB_{i}\left( t\right) dB_{j}\left( t\right) ^{\ast }=\delta _{ij}\,dt.
\label{eq:QIT}
\end{eqnarray}
Taking the Fock vacuum state, we then have \cite{HP,partha, ALV}
\begin{eqnarray}
\textrm{tr}\left\{ \varrho _{\text{sys}}\,e^{t\mathcal{L}}\left( X\right)
\right\} &\equiv& \text{tr}\big\{ \varrho _{\text{sys}}\,\otimes | \Omega \rangle \langle \Omega | \nonumber \\
&& U\left( t,0\right) ^{\ast }\left( X\otimes I_{\mathfrak{F}}\right) U\left(
t,0\right) \big\} ,
\label{eq:Fock_expect}
\end{eqnarray}
with $\mathcal{L}$ given as in (\ref{eq:GKS_Lindblad}).

\subsection{Applying DD to the System}
We similarly apply the dynamical decoupling scheme as before, with $\tilde{U}_{\mathbf{v}}\left( t,s\right) $ as given by (\ref{eq:SQT_DD_step}), but now with 
\begin{eqnarray}
\tilde{U}_{v_{k}}\left( t_{k+1},t_{k}\right) =\left( v_{k}\otimes I_{%
\mathfrak{F}}\right) ^{\ast }U\left( t_{k+1},t_{k}\right) \,\left(
v_{k}\otimes I_{\mathfrak{F}}\right) .
\label{eq:AQT_DD_step}
\end{eqnarray}
Note that the decoupling scheme is only applied to the system!

Our aim is to apply the decoupling scheme with limiting small steps $\left(\tau \rightarrow 0\right) $. To obtain a continuous limit description of the dynamics we have to work in the regime where the dynamical decoupling is both effective and may be treated as an infinitesimal - in other words $T_{dd}$ is small, with $\tau $ proportionately smaller. At this level of approximation, we may approximate the unitaries as 
\begin{eqnarray}
\lefteqn{\tilde{U}\left(
t_{k+1},t_{k}\right)} \notag\\
&=&I_{\text{sys}}\otimes I_{\mathfrak{F}}-v_k^\ast \left( \sum_{j=1}^{n} \frac{1}{2} L_{j}^{\ast
}L_{j}+iH\right) v_k \otimes \tau   \nonumber \\
&&+\sum_{j=1}^{n}v_k^\ast L_{j}v_k\otimes \left[ B_{j}\left( t_{k+1}\right) -B_{j}\left(
t_{k}\right) \right] ^{\ast }  \nonumber \\
&&-\sum_{j=1}^{n} v_k^\ast L_{j}^{\ast } v_k \otimes \left[ B_{j}\left( t_{k+1}\right)
-B_{j}\left( t_{k}\right) \right] .  \label{eq:QSDE_U_approx}
\end{eqnarray}

Let us now define the conditional expectation $\mathbb{E}_{\text{vac}}$ from the system+noise algebra operators onto the system operators by tracing out the noise in the vacuum state, that is 
\begin{eqnarray*}
\mathrm{tr}_{\text{sys}}\left\{ \rho \mathbb{E}_{\text{vac}}\left[ Z\right]
\right\} \equiv \mathrm{tr}_{\text{sys+noise}}\left\{ \rho \otimes |\Omega \rangle
\langle \Omega |\,Z\right\}
\end{eqnarray*}
for every density matrix $\rho $ of the system and operator $Z$ of the system+noise.

Using the approximation (\ref{eq:QSDE_U_approx}) followed by (\ref{eq:LLN}), we see that $\mathbb{E}_{\text{vac}}\left[ \tilde{U}_{\mathbf{v}}\left( t_{0}+T_{dd},t_{0}\right) \right] $ is, to leading order, 
\begin{eqnarray*}
&\approx &I_{\text{sys}}-\tau \sum_{k=1}^{N_{dd}}v_{k}^{\ast }\left(
\sum_{j=1}^{n} \frac{1}{2} L_{j}^{\ast }L_{j}+iH\right) v_{k} \\
&\approx &\left\{ 1-\lambda\, T_{dd}  \right\} I_{\text{sys}},
\end{eqnarray*}
where
\begin{eqnarray}
\lambda =\frac{1}{d} \mathrm{tr}\left(
\sum_{j=1}^{n} \frac{1}{2} L_{j}^{\ast }L_{j}+iH\right).
\end{eqnarray}
With the understanding that $T_{dd}$ is infinitesimal, the flow property implies that 
\begin{eqnarray*}
\mathbb{E}_{\text{vac}}\left[ \tilde{U}_{\mathbf{v}}\left( t_{0}+T,t_{0}\right) %
\right] =e^{-T\lambda } \, I_{\text{sys}}
\end{eqnarray*}
for finite $T$.

Our interest is in the Heisenberg dynamical flow. For $X$ a fixed system operator, we are interested in 
\[
\mathbb{E}_{\text{vac}}\left[ \tilde{U}_{\mathbf{v}}\left(
t_{0}+T_{dd},t_{0}\right) ^{\ast }\left( X\otimes I_{\text{noise}}\right) 
\tilde{U}_{\mathbf{v}}\left( t_{0}+T_{dd},t_{0}\right) \right] 
\]
and to the same level of approximation this equals 
\begin{eqnarray*}
&\approx &X-\tau \sum_{k=1}^{N_{dd}}v_{k}^{\ast }\left(
\sum_{j=1}^{n} \frac{1}{2} L_{j}^{\ast }L_{j}-iH\right) v_{k}X \\
&&-\tau \sum_{k=1}^{N_{dd}}Xv_{k}^{\ast }\left( \sum_{j=1}^{n} \frac{1}{2} L_{j}^{\ast
}L_{j}+iH\right) v_{k} \\
&&+\tau \sum_{k=1}^{N_{dd}}\sum_{j=1}^{n}v_{k}^{\ast }L_{j}^{\ast
}v_{k}Xv_{k}^{\ast }L_{j}v_{k} \\
&\approx &X-\frac{1}{d}T_{dd}\sum_{j=1}^{n}\mathrm{tr}\left( L_{j}^{\ast
}L_{j}\right) X \\
&&+T_{dd}\,\frac{1}{\#\mathscr{V}}\sum_{v\in \mathscr{V}%
}\sum_{j=1}^{n}v^{\ast }L_{j}^{\ast }v \, X \, v^{\ast }L_{j}v,
\end{eqnarray*}
where we have used the incremental form of (\ref{eq:QIT}). The final answer may then be written as 
$X+\mathcal{\bar{L}}\left( X\right) T_{dd}$ where we introduce the super-operator
\begin{eqnarray}
\mathcal{\bar{L}}\left( X\right)  &=&\frac{1}{\#\mathscr{V}}\sum_{v\in %
\mathscr{V}}\sum_{j=1}^{n}v ^{\ast }L_{j}^{\ast }v Xv ^{\ast
}L_{j}v  \nonumber \\
&&-\frac{1}{d}\sum_{j=1}^{n}\mathrm{tr}\left( L_{j}^{\ast }L_{j}\right) X.
\end{eqnarray}
We note that $\mathcal{\bar{L}}$ is a GKS-Lindblad generator as is readily seen by writing it in the standard form
\begin{eqnarray}
\lefteqn{\mathcal{\bar{L}}\left( X\right)} \notag\\  
&=&\sum_{v\in \mathscr{V}}\sum_{j=1}^{n}%
\frac{1}{2}\bigg\{ \left[ R_{v,j}^{\ast },X\right] R_{v,j} 
+R_{v,j}^{\ast }\left[ X,R_{v,j}\right] \bigg\} , \qquad
\label{eq:standard_Lindblad_form}
\end{eqnarray}
where 
\begin{eqnarray}
R_{v,j}=\frac{1}{\sqrt{\#\mathscr{V}}}v ^{\ast }L_{j}v .
\label{eq:Rs}
\end{eqnarray}

Again we have been operating under the assumption that $T_{dd}$ is infinitesimal. For finite times $T$ we have 
\begin{eqnarray}
&&\mathbb{E}_{\text{vac}}\left[ \tilde{U}_{\mathbf{v}}\left( t_{0}+T,t_{0}\right)
^{\ast }\left( X\otimes I_{\mathfrak{F}}\right) \tilde{U}_{\mathbf{v}}\left(
t_{0}+T,t_{0}\right) \right] \nonumber \\
&\equiv & e^{T\mathcal{\bar{L}}}\left( X\right) .  \label{eq:limit_QDS}
\end{eqnarray}
Therefore, we see that we do not succeed in dynamically decoupling our system - this is clear from the fact that we obtain the genuinely
dissipative GKS-Lindblad generator $\bar{\mathcal{L}}$.

We remark that the dissipation associated with $\mathcal{\bar{L}}$ is now
\begin{eqnarray}
D_{\mathcal{\bar{L}}} \left( X,X \right)  &=& \frac{1}{\# \mathscr{V}} \sum_{v\in \mathscr{V}}\sum_{j=1}^{n}%
  \left[ X, v^\ast L_j v\right]^\ast   \left[ X, v^\ast L_j v\right] \nonumber \\
	&=& \frac{1}{\# \mathscr{V}} \sum_{v\in \mathscr{V}} 
v^\ast  D_{\mathcal{L}} \left( v^\ast  Xv, v^\ast Xv \right) v ,
\end{eqnarray}
which is typically different from that the original $\mathcal{L}$. This opens up the possibility that the dynamical decoupling may result in extending the dissipation to observables that were previously invariant under the original dynamics.

\subsection{Examples}
For example, in the dephasing case we have a single coupling operator $L= \sqrt{\gamma} \sigma_z$ and in this case, for
the decoupling scheme $\mathscr{V} = \{ I ,\sigma_x \}$ which worked in the shallow pocket model, we have
$\mathcal{\bar{L}} \equiv \mathcal{L} $, so despite the DD scheme the generator remains unchanged. This also remains true 
for the case of the Pauli group $\mathscr{V}_{\text{Pauli}} = \{ \pm I ,\pm \sigma_x , \pm \sigma_y , \pm \sigma_z \}$.

\bigskip

In the case of damping, we take $L = \sqrt{\gamma} \sigma_-$ which has generator $\mathcal{L}_- X = \gamma \sigma_+ X \sigma_-
-\frac{1}{2} \gamma X \sigma_+ \sigma_-  -\frac{1}{2} \gamma \sigma_+ \sigma_-  X$. Without dynamical decoupling, this would imply a master equation where the state converges to the ground state (south pole of the Bloch sphere). If we applied DD with Pauli group $\mathscr{V}_{\text{Pauli}}$, then we would in fact find $\mathcal{\bar{L}} \equiv  \frac{1}{2} \mathcal{L}_+ + \frac{1}{2} \mathcal{L}_-$, where $ \mathcal{L}_+$ is the
generator corresponding to $L = \sqrt{\gamma} \sigma_+$. This is because conjugation with either $\pm I$ or $\pm \sigma_z$ change $\sigma_-$ to $\pm \sigma_-$, while conjugation with either $\pm \sigma_x$ or $\pm \sigma_y$ change $\sigma_-$ to $\pm \sigma_+$. 
This DD therefore results in the convergence to a different state - the minimum information state (centre of the Bloch sphere). The damping rates are same in both cases: the decaying terms have factors $e^{-\gamma t} $ on the diagonal, and $e^{-\gamma t /2}$ off-diagonal. Arguably the DD scheme is making things worse in this case.

\subsection{Remarks}
It should be pointed out that the flow $U(t,0)$ maps from a co-cycle with respect to the time-shift unitary on $\mathfrak{F}$. Specifically, let $V^0(t)$ be the unitary shifting vectors on $\mathfrak{F}$ by time $t$ then
\begin{eqnarray}
U(t+s,0) \equiv V^0(s)^\ast U(t,0) V^0 (s) \, U(s,0) .
\end{eqnarray}
However $U(t,0)$ is strongly continuous in $t$ (though obviously not differentiable) and the unitaries defined
by
\begin{eqnarray}
V(t) \triangleq V^0 (t) \, U(t,0)
\end{eqnarray}
then defines a strongly continuous one-parameter (we have to extend $t$ to negative times!) group and so possesses a Stone generator $H_{\text{C.G.}}$
which we call the Chebotarev-Gregoratti Hamiltonian:
\begin{eqnarray}
V(t) \equiv e^{-it H_{\text{C.G.}}}.
\end{eqnarray}
As the shift acts only on the Fock space, it $V^0(t)$ leaves the system space invariant, so therefore
\begin{eqnarray}
U(t,0)^\ast (X \otimes I_{\mathfrak{F}} ) U(t,0) \equiv V(t)^\ast (X \otimes I_{\mathfrak{F} }) V(t) .
\end{eqnarray}
In particular, the fact that we have used a cocycle rather than a group in (\ref{eq:Fock_expect}) is not important. The cocycle property however does make a difference for multi-time time-ordered correlation kernels and this is where the Markov nature comes in - in particular, we obtain the correct kernels associated with $\Phi_t = e^{t \mathcal{L}}$, with $\mathcal{L}$ given by (\ref{eq:GKS_Lindblad}), as stipulated by (\ref{eq:multitime}).

We now have
\begin{eqnarray}
\lefteqn{\tilde{U}_{\mathbf{v}}\left( t,s\right)} \\
&=&
\tilde{U}_{v_N} \left( t_{N},t_{N-1}\right) \cdots \tilde{U}_{v_2} \left( t_{2},t_{1}\right) \tilde{U}_{v_1} \left( t_{1},t_{0}\right) \nonumber \\
&=& V^0(t) \,\, \tilde{V}_{v_N}\left( t_{N}-t_{N-1}\right) \cdots \tilde{V}_{v_2} \left( t_{2}-t_{1}\right) \tilde{V}_{v_1}  ( t_{1}) ,\nonumber \\
\label{eq:new} 
\end{eqnarray}
where now
\begin{eqnarray}
\tilde{V}_{v_{k}} ( t ) = (v_{k} \otimes  I_{\mathfrak{F}}) ^{\ast }V(t)  (v_{k} \otimes  I_{\mathfrak{F}}) .
 \label{eq:V_DD_step}
\end{eqnarray}
This is, of course, consistent with the dynamical decoupling scheme for Hamiltonian evolution coupling the system to the Fock bath via the
Chebotarev-Gregoratti Hamiltonian $H_{\text{C.G.}}$.

Our results from the previous section would therefore show that the Chebotarev-Gregoratti Hamiltonian $H_{\text{C.G.}}$ cannot be dynamically decoupled 
using schemes acting solely on the system space.

\addtolength{\textheight}{-3.6cm}

\section{Discussion}
In this paper we have studied dynamical decoupling associated with dilations of quantum dynamical semigroups. A fixed semigroup in general will possess several inequivalent dilations, and we have seen instances where a dissipative semigroup has a dilation where dynamical decoupling works, and another where it does not. 

To better understand this, it is worth recalling that there are chaotic hierarchies of noise models. This was first articulated by Accardi \cite{Accardi90} for quantum systems: at the top are independent increment processes; then Markov processes with an expected past filtration; and so on down to processes with deterministic algebraic filtration.

If we look at the minimal stochastic model (coupling our system to an unstable particle), then the \lq\lq noise\rq\rq\, is effectively just a single real random variable $\Lambda$. This type of model would lie close to the bottom of the Accardi hierarchy of quantum chaoticity. Effectively it is just a Hamiltonian randomized by a single random variable (which we could just as easily take to be classical).

The quantum stochastic models considered in Section \ref{sec:QSDE_models} in contrast requires a Fock space (Bose second quantization) of time-pulse wavefunctions. We are now at the maximum end of chaoticity as quantum noise goes. This level is necessary if we are to realize Markovianity.

A well known rule of thumb is that it is impossible to dynamically decouple a time-dependent Hamiltonian evolution if the timescales at which the Hamiltonian changes are shorter than the ones to which we go with our periodic dynamical decoupling modulations. Our results for open systems are in line with this - for models with high chaoticity in the noise, the timescale at which the dynamics changes is small (effectively zero for quantum stochastic models), and so dynamical decoupling turns out to be limited
(complete DD being impossible for quantum stochastic models even in the limit of infinite rapidity).

In \cite{A_15} it is argued that the ability to dynamically decouple should be a mechanism to distinguish standard quantum theories (i.e., governed by a fixed Hamiltonian) from what they term \lq\lq alternative quantum theories\rq\rq . For some reason, they characterize quantum stochastic models as alternative quantum theories, even though they are just cocycles resulting from going over to the interaction picture for the evolution under the Chebotarev-Gregoratti Hamiltonian relative to the evolution under the Fock space shift dynamics - granted though that this is a singular perturbation. A more workable statement would be to say that a given quantum dynamical semigroup (with dissipative GKS-Lindblad generator $\mathcal{L}$) can have different dilations: all we ask is that the dilation gets the one-point correlations (i.e., the means) correct without requirements on the higher multi-time correlation kernels. Whether we may dynamically decouple a specific dilation is not a property of the semigroup, or equivalently of its generator $\mathcal{L}$, but instead a property of the degree of chaoticity of noise we utilize in the dilation. As such, one cannot tell if an system can be dynamically decoupled, or not, just by looking at the master equation. One also needs to know the multi-time correlations to the environment. This does however raise an interesting question, namely is there an identifiable level in the hierarchy of model chaoticity below which dynamical coupling will work, but above which it will not?

\section{Conclusion}
The implications of this paper are that \textit{complete} dynamical decoupling does not work for quantum stochastic evolutions. 
To be clear, the goals of DD are more realistic than complete decoupling in practical situations, often seeking to slow down the
rate of decoherence, and maybe only for a particular subalgebra of operators. Our results however are general enough: the question of how applicable they are ultimately comes down to modelling issues. If one is working with these type of models, which is typically the case in quantum optics situations, then one must be working at a time scale where the bath auto-correlation time is negligible - and the while the dynamical decoupling may be considered as periodic and rapid, the time steps are nevertheless large compared to the bath auto-correlation time. This sets the various time scale regimes. In principle, as the decoupling operations are applied at longer
time scales that the bath auto-correlation, we should not be too surprised to see that the dynamical decoupling is no longer effective.

\section*{Acknowledgement}
We acknowledge discussions with Daniel Burgarth.

\bibliographystyle{ieeetran}

\end{document}